\documentclass[a4paper,11pt]{article}
\usepackage{amssymb,amsmath, amsthm}
\usepackage{epsfig}
\usepackage[english]{babel}
\usepackage{latexsym}
\usepackage{amscd}
\usepackage{braket}
\usepackage{amssymb}
\usepackage{slashed}
\usepackage{lmodern,textcomp}
\newcommand{\splitcell}[2][c]{%
	\begin{tabular}[c]{@{}c@{}}\strut#2\strut\end{tabular}%
}

\begin{document}

\title{A path integral based model\\for stocks and order dynamics}
\date{}
\author{Giovanni Paolinelli \and Gianni Arioli}
\maketitle

\begin{center}
{\small Dipartimento di Matematica}\\
{\small Politecnico di Milano }\\
{\small piazza Leonardo da Vinci 32, 20133 Milano, Italy}\\
\texttt{\small giovanni.paolinelli@polimi.it, gianni.arioli@polimi.it}
\par\end{center}{\small \par}

\textbf{Keywords}:  Stock prices, econophysics, path integral, gauge theory, financial markets.

\begin{abstract}
We introduce a model for the short-term dynamics of financial assets based on an application to finance of quantum gauge theory, developing ideas of Ilinski.
We present a numerical algorithm for the computation of the probability distribution of prices and compare the results with {\em APPLE} stocks prices and the {\em S\&P500} index.
\end{abstract}

\section{Introduction}
In \cite{Il} Ilinski introduced a model for the short-term dynamics of stocks and orders based on a gauge interpretation of classical finance. A similar approach has been developed in \cite{D-F,D-F2,D-F3}, where 
a model based on quantum field theory has been developed.
The approach to the financial mathematical problems with the quantum mechanics is called quantum finance. In the last 20 years, the path integral approach to quantum mechanics introduced by Feynman \cite{Fey1} has turned out to be particularly suitable for financial applications, see also \cite{Ba,Ba2,MNM,MMNAF,BMMN,DLT}.
We further analyse and extend these ideas and develop a model that provides results in good agreement with real market data. 

Ilinski's model describes the amount of cash and share in the portfolio and all possible trading configurations, and in particular the impact of the orders in the stocks dynamics. Quantum mechanics plays a fundamental role in providing a robust mathematical background to describe the fundamental ideas of this theory. In particular, the discrete nature of the portfolio, characterized by an integer number of stock and cash is modelled by the coherent-state path-integral. We generalize this model and introduce an algorithm to compute the coherent-state path-integral proposed by Ilinski.

The model is tested against data of {\em APPLE} stocks and the {\em S\&P500} index with a time step of one minute; the agreement between real data and the model covers four orders of magnitude. In particular we get a good fit also of the fat-tails.

It is interesting to observe that the fat tails phenomenon can be seen as an effect of the orders. We point out that the shape of the tails has a significant impact on the risk-free rate. More precisely, in \cite{P-D} it is shown that the rate of return of an investment with no risk of financial loss and the term premium, i.e. the compensation that investors require for bearing the risk that short-term Treasury yields do not evolve as they expected, are miscomputed if the fat tails are ignored.

We show a direct relationship between the kurtosis of the PDF and the strength of the perturbation caused by the orders; in fact, the model without orders provides results equivalent to the Geometric Brownian Motion.

Our model entails five parameters; the same amount of the models in \cite{D-F} and \cite{Il}. We analyse the impact of the parameters on the PDF and provide a financial interpretation.

\section{A simpler model}

We present in this section the basic concepts of Ilinski's theory; the reader interested in a detailed explanation should refer to \cite{Ila} and \cite{Il}.

Ilinski's starting point is the basic idea that it is not possible to earn money without risk; if this were the case, then we would have an arbitrage opportunity. Consider  an elementary market model where it is possible to buy a non-risky asset $B$ and a risky asset $S$ with the same initial value. If we assume temporarily  that $S$ is not a risky asset, then after a time $T$ the final values of $B$ and $S$ must to be equal. Otherwise it would be possible to perform an arbitrage operation; that is, we could borrow and sell the under-performing asset, and then use the capital obtained to buy the over-performing one. When we sell the over-performing asset we have enough money to buy back the under-performing asset and also have some money left, the arbitrage revenue. This situation does not occur in the reality since we cannot know the value at a future time of $S$, i.e., we do not know in advance whether $S$ is the under- or the over-performing asset. This implies that a revenue from the previous situation can only be achieved with some amount of risk.

In $\cite{Il}$ Ilinski shows a strategy involving stocks and cash which generates a positive revenue independently of the final value of the risky asset $S$. Such strategy is called {\em arbitrage}. In classical finance such amount is assumed to be zero. Ilinski proposes a weaker assumption, which entails a minimization of the arbitrage.

Denote by $A_g(\{S\})$ the gain of the arbitrage described above; in Ilinski's model this quantity is treated as the action in a Lagrangian description of the dynamics of the system. We assume that the probability associated to $A_g(\{S\})$ is given by
\begin{equation}
P(\{S\})=Ne^{-\beta A_g(\{S\})}.
\label{eq:arbi}
\end{equation}     
This assumption represents the main link with the quantum mechanics and with path integrals in particular.
The core idea in the path integral representation of quantum mechanics consists in the fact that all  trajectories are considered, but those achieving a lower action are more probable.
In Ilinski's financial model the least action ($\hbar\to0$ in quantum mechanics) corresponds to zero arbitrage. Trajectories with small arbitrage (quantum fluctuations) are considered, but their probability decreases when they get farther apart from the zero arbitrage trajectory. The role of Plank's constant is played by the variance of the financial asset.

We also assume that the stock dynamics is invariant under the change of the currency. This concept is well known in physics as gauge-invariance. We require our theory not to depend of the choice of the {\em numeraire} for any asset at any moment of the time. Agents do not start to behave in a different way  because they are dealing with 100 pence instead of 1 $\pounds$ or the equivalent amount of money in $\$$.
This invariance must be encoded in all quantities described in this theory. We assume that the probability of a certain amount of arbitrage $A_g(\{S\})$ does not change if the assets are expressed in a different currency.\par 
In physics, the equations that describe the system in a gauge theory are invariant under the transformation induced by the action of a group, either local or global. If we want to denote 10$\$$ in $\pounds$, i.e, perform a change of gauge, we have to multiply the capital by a positive number which is the conversion ratio; the previous number is an element of the gauge group. This is true for all possible currency conversions, therefore the gauge group in this context is the multiplicative
$\mathbb{R}^+$.

Ilinski uses the previous framework to obtain the conditional probability for the stock price. He considers a discrete time model, where time takes the values $t_i=i\Delta$ for some $\Delta>0$, and then the price of an asset is denoted by $S_i=S(t_i)=S(i\Delta)$, $i=0,\ldots,N$. Denote by $S_0$ the price at time $0$ and $S$ the final price at time $T=N\Delta$; Ilinski proves that:

\begin{equation}
P(T,S|0,S_0)=
\left(\prod_{i=1}^{N-1}\int_0^\infty\frac{dS_i}{S_i}\right)\label{prob}
\exp\left(\frac{-1}{2\Delta\sigma^2}\sum_{i=0}^{N-1}R_i\right)\,,
\end{equation}
where
$$
R_i=S^{-1}_ie^{r_2\Delta}S_{i+1}e^{-r_1\Delta}+S_ie^{-r_2\Delta}S^{-1}_{i+1}e^{r_1\Delta}-2
$$
is the revenue at time $t_i$ of the double arbitrage and the constants $r_1$ and $r_2$ represent the interest rates respectively of the cash and of the risky asset. The  exponential is the probability described in $(\ref{eq:arbi})$, whereas  $\sigma$ is the variance of the prices of the risky asset. Note that the values $R_i$ are gauge invariant.

\begin{figure}[ht]
\begin{center}
		\includegraphics[width=8cm]{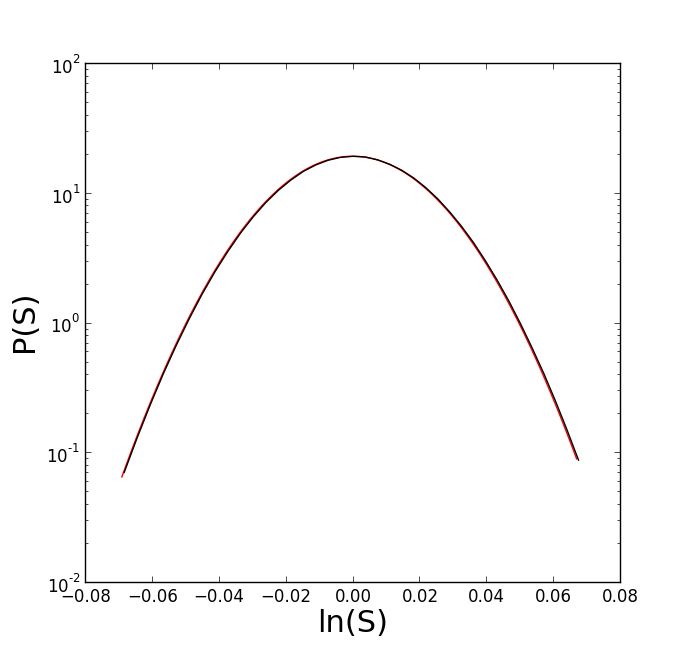}
	\end{center}
\caption{Probability density functions with respect to the final stock price $S$, in logarithmic scale. Obtained by  numerical computation of (\ref{prob}) --red-- and  with the analytic formula --black--. Both curves have  identical parameters  $\sigma=0.00648$, $T=10$, $N=10$ and $r_1=r_2=0$. }
\label{fig:gbm}
\end{figure}

The product of differentials is the path space differential used to sum all over the possible trajectories from $S(0)$ to $S(T)$; details about this concept can be found in \cite{Fey}. The measure is the gauge invariant  expression $dS_i/S_i$.
In \cite{Il}, Ilinski proves that (\ref{prob}) results in a normal PDF. Figure \ref{fig:gbm} shows the result of a numerical simulation that corroborates such result. 

\section{The model with orders}

In \cite{Il} Ilinski also introduces a generalization of the previous model consisting in the addition of a perturbation which takes into account the effect of the orders in the stocks dynamics. In this paper we show that a modified perturbation generates a leptokurtic probability distribution of the returns.

Ilinski adds to the action a term which describes the dynamics of the orders, so that the price goes up (down) when somebody buys (sells) the stock. The action is given by
\begin{equation}\label{gpot}-\beta\mathcal{A}_g(S)=-\frac{1}{2\sigma^2\Delta}\sum_{i=0}^{N-1}\bigg(\log(S_{i+1})-\log(S_{i})-\mu\Delta-\frac{N_i}{\lambda}\bigg)^2\,;\end{equation}
The derivation can be found in \cite{Il}.
 
The terms $N_i$ represent the net amount of the orders at time $t_i$, its value is positive if the net amount is a buy order, negative otherwise, while $\lambda$ represents the share liquidity.\par
The initial allocation of the portfolio is described by the pair $(n_1,m_1)$ which represents the amounts of cash and  share at the initial time; the final allocation is denoted by $(n,m)$.
Because of the gauge invariance, we can express the share and money values in the same unit, so that a unit of cash can be traded for a share.\par
We assume the closed environment hypothesis, which implies a constant amount of lots at all  times;
denoting by $M$ the total number of traded lots (both money and shares) we have  $n_1+m_1=n+m=M$. In this context the initial configuration of the system is the capital allocation $(n_1,m_1)$; at each time step $i$, the system  evolves to the capital allocation $(n_i,m_i)$ up to time $T$, where it achieves the final allocation $(n,m)$. In order to consider the effect of the orders on the price dynamics, it is necessary to compute all the possible paths in the capital allocation space and add the effect of each path.

This computation is achieved with the Coherent State Path Integral (CSPI), see \cite{Il,Fey,Edu}. The numbers $(n_i,m_i)$ are integers, and in the context of the CSPI they are described as 
\[n_i=\bar{\psi}_{1,i}\psi_{1,i},\]
\[m_i=\bar{\psi}_{2,i}\psi_{2,i},\] 
where $\{\psi_{j,k}\}$ are complex numbers, corresponding to the creation/annihilation operators.\par
We  denote an order $N_i$ via the $\psi_{j,k}$ operators. Buying $k$ stocks we lose  $k$ units of cash, and vice versa. We have: 
\[N_i=\delta_i [\bar{\psi}_{2,i}\psi_{2,i}-\bar{\psi}_{1,i}\psi_{1,i}].\]
Where the symbol $\delta_i$ stands for the forward different quotient, i.e.,
\[\delta_i h(i)=\frac{h(i+1)-h(i)}{\Delta},\]
with $\Delta$ equal to the minimal time frame.\\
The dynamics of the variables $\bar{\psi},\psi$ is described by a Hamiltonian
\[H(t)(\bar{\psi},\psi)=H(i)_{jk}\bar{\psi}_{j,i+1}\psi_{k,i},\] which links $\psi_{(j,k),i}$ to $S(i\Delta)=S_i$. The following expression is derived in \cite{Il}:
\begin{equation}\label{ham}
H(i)_{jk}=
\begin{bmatrix}
0       & \gamma S^{\tilde{\beta}}_ie^{-\tilde{\beta}r_1\Delta}e^{-\tilde{\beta}\mu t} \\
\gamma S^{-\tilde{\beta}}_ie^{-\tilde{\beta}r_2\Delta}e^{\tilde{\beta}\mu t}       & 0  \\
\end{bmatrix},
\end{equation}
\[\gamma=(1-tc)/\Delta'.\]
Here $tc$ is the relative cost of the transaction; $\Delta'$ is the time step of the model in the Hamiltonian dynamics and $\tilde{\beta}$ denotes the amplitude of the price variations.

In our simulations we assume $\mu=r_1=r_2=tc=0$ and $\Delta'=\Delta$.
Given the Hamiltonian, we compute the propagator:
\begin{align}\label{prop} 
&\bra{\bar{\psi}_{1,N},...,\bar{\psi}_{Z,N}}|U(T=N\Delta,0)|\ket{\psi_{1,0},...,\psi_{Z,0}}=\nonumber\\
&\prod_{j=1}^{N-1}\prod_{k=1}^{Z}\int\frac{d\psi_{k,j}d\bar{\psi}_{k,j}}{2i\pi}\times
\exp\bigg[-\sum_{j=1}^{N-1}\sum_{k=1}^{Z}\psi_{k,j}\bar{\psi}_{k,j}+\sum_{j=1}^{N-1}\sum_{k=1}^{Z}\psi_{k,j}\bar{\psi}_{k,j+1}\bigg]\times\\
&\exp\bigg[\Delta\sum_{j,k=1}^{Z}H(N-1)_{jk}\bar{\psi}_{i,N}\psi_{k,N-1}+\ldots
+\Delta\sum_{j,k=1}^{Z}H(0)_{jk}\bar{\psi}_{i,1}\psi_{k,0}\bigg]\,.\nonumber
\end{align}
The quantity inside the first square bracket corresponds to 
\begin{equation}\label{sumtrik}\sum_{k=1}^{Z}\bar{\psi}_{k,0}\psi_{k,0}+\sum_{k=1}^{Z}\sum_{j=0}^{N-1}(\bar{\psi}_{k,j+1}-\bar{\psi}_{k,j})\psi_{k,j}.\end{equation}
Substituting $(\ref{ham})$ and $(\ref{sumtrik})$ in $(\ref{prop})$ we obtain:

\begin{align}\label{secprop} 
&\bra{\bar{\psi}_N}\hat{U}(T=N\Delta ,0)\ket{\psi_0}=e^{\bar{\psi}_{1,0}\psi_{1,0}+\bar{\psi}_{2,0}\psi_{2,0}}\int\prod_{k=1,2}\prod_{i=1}^{N-1}\frac{d\psi_{k,j}d\bar{\psi}_{k,j}}{2i\pi}\times\nonumber\\
&\exp\bigg[\sum_{i=0}^{N-1}(\bar{\psi}_{1,i+1}\psi_{1,i}-\bar{\psi}_{1,i}\psi_{1,i}+\bar{\psi}_{2,i+1}\psi_{2,i}-\bar{\psi}_{2,i}\psi_{2,i}+S_i^{\tilde{\beta}}\bar{\psi}_{1,i+1}\psi_{2,i} +S_i^{-\tilde{\beta}}\bar{\psi}_{2,i+1}\psi_{1,i} )\bigg]\,.
\end{align}
We introduce the hydrodynamical variables
 \begin{equation}\label{hydro}\bar{\psi}_k=\sqrt{M\rho_k}e^{i\phi_k}\quad\quad\psi_k=\sqrt{M\rho_k}e^{-i\phi_k}\,,\end{equation}
where $\rho \in [0,1]$ and $\phi \in [0,2\pi]$.
Note that, because of the close environment  assumption,
$M(\rho_1+\rho_{2})=M$ and then $\rho_1=1-\rho_2$.

We write $(\ref{secprop})$ in the hydrodynamical variables.
Starting from
\begin{equation}
\bar{\psi}_{1,i+1}\psi_{1,i}-\bar{\psi}_{1,i}\psi_{1,i}=[\sqrt{\rho_{1,i+1}}e^{i(\phi_{1,i+1}-\phi_{1,i})}-\sqrt{\rho_{1,i}}]\sqrt{\rho_{1,i}}\,,\label{eq:hdv}
\end{equation}
if we assume
\[e^{i(\phi_{1,i+1}-\phi_{1,i})}\simeq1+i(\phi_{1,i+1}-\phi_{1,i})\]
then (\ref{eq:hdv}) becomes:
\[\bar{\psi}_{1,i+1}\psi_{1,i}-\bar{\psi}_{1,i}\psi_{1,i}=[\sqrt{\rho_{1,i+1}}-\sqrt{\rho_{1,i}}+i\sqrt{\rho_{1,i+1}}(\phi_{1,i+1}-\phi_{1,i})]\sqrt{\rho_{1,i}};\]
recalling the definition of  the forward  difference quotient, we obtain:
\[\bar{\psi}_{1,i+1}\psi_{1,i}-\bar{\psi}_{1,i}\psi_{1,i}=\Delta\sqrt{\rho_{1,i}}(\delta_i \sqrt{\rho_{1,i}})+i\Delta\sqrt{\rho_{1,i+1}\rho_{1,i}}(\delta_i\phi_{1,i});\]
recalling that \[\sqrt{\rho_{1,i}}(\delta_i\sqrt{\rho_{1,i}})=\frac12\delta_i\rho_{1,i},\] 
we get
\[\bar{\psi}_{1,i+1}\psi_{1,i}-\bar{\psi}_{1,i}\psi_{1,i}+\bar{\psi}_{2,i+1}\psi_{2,i}-\bar{\psi}_{2,i}\psi_{2,i}=\]\begin{equation}\label{firstpart}=i\bigg[(\phi_{1,i+1}-\phi_{1,i})\sqrt{\rho_{i+1}\rho_{i}}+(\phi_{2,i+1}-\phi_{2,i})\sqrt{(1-\rho_{i+1})(1-\rho_{i})}\bigg],\end{equation}
where the conservation law $\rho_1+\rho_2=1$ has been used.

We also have
\[S_i^{\tilde{\beta}}\bar{\psi}_{1,i+1}\psi_{2,i}+S_i^{-\tilde{\beta}}\bar{\psi}_{2,i+1}\psi_{1,i}=\]\begin{equation}\label{secondpart}=S_i^{\tilde{\beta}}\sqrt{\rho_{i+1}(1-\rho_{i})}e^{i(\phi_{1,i+1}-\phi_{2,i})}+S_i^{-\tilde{\beta}}\sqrt{(1-\rho_{i+1})\rho_{i}}e^{i(\phi_{2,i+1}-\phi_{1,i})}.\end{equation}
Using the equation $(\ref{firstpart})$ and $(\ref{secondpart})$ in $(\ref{secprop})$ we obtain the formula for the propagator: 
\[\bra{\bar{\psi}_N}\hat{U}(T=N\Delta ,0)\ket{\psi_0}=e^{M}\prod_{k=1,2}\prod_{i=1}^{N-1}\int_{0}^{1}d\rho_{k,i}\frac{\rho_{k,i}}{\pi}\int_{0}^{2\pi}d\phi_{k,i} \]\[\times \exp\bigg[M\sum_{i=0}^{N-1}\bigg(i\{(\phi_{1,i+1}-\phi_{1,i})\sqrt{\rho_{1,i+1}\rho_{1,i}}+(\phi_{2,i+1}-\phi_{2,i})\sqrt{\rho_{2,i+1}\rho_{2,i}}\}+\]\[+S_i^{\tilde{\beta}}\sqrt{\rho_{1,i+1}\rho_{2,i}}e^{i(\phi_{1,i+1}-\phi_{2,i})}+S_i^{-\tilde{\beta}}\sqrt{\rho_{2,i+1}\rho_{1,i}}e^{i(\phi_{2,i+1}-\phi_{1,i})}\bigg)\bigg].\]
The previous propagator depends on all possible paths in the portfolio space; in order to obtain the conditional probability we need to link it with the price dynamics using $(\ref{gpot})$.
We can write $N_i$ as:
\begin{equation}
N_i=\delta_i [\bar{\psi}_{2,i}\psi_{2,i}-\bar{\psi}_{1,i}\psi_{1,i}]=\delta_i [\rho_1-\rho_2]=2\delta_i\rho_i\,,
\end{equation}
so that equation $(\ref{gpot})$ becomes
\[-\beta\mathcal{A}_g(S)=\sum_{i=0}^{N-1}\frac{-1}{2\sigma^2\Delta}\bigg(\log(S_{i+1})-\log(S_{i})-2\alpha[(\rho_{i+1}-\rho_{i})]\bigg)^2 \quad  \textup{with}\quad\alpha=M/\lambda.\]
We consider now the propagator above together with the stock price action $(\ref{gpot})$, and we obtain the new propagator which takes into account both the trajectories in the spaces of portfolio allocation and the stock prices.
\[\]
\[\bra{\bar{\psi_N}}\hat{U}(T=N\Delta ,0)\ket{\psi_0}=\]\[e^{M}\prod_{k=1,2}\prod_{i=1}^{N-1}\int_{0}^{1}d\rho_{i}\frac{\rho_{i}}{\pi}\int_{0}^{2\pi}d\phi_{k,i}\int_{-\infty}^{\infty}d\log(S_i)\times \exp[-\beta\mathcal{A}_g(S)+\tilde{S}].\]
with:
\[-\beta\mathcal{A}_g(S)=\sum_{i=0}^{N-1}\frac{-1}{2\sigma^2\Delta}\bigg(\log(S_{i+1})-\log(S_{i})-2\alpha[(\rho_{i+1}-\rho_{i})]\bigg)^2\]
\[\tilde{S}=M\sum_{i=0}^{N-1}\bigg(\quad i\bigg[(\phi_{1,i+1}-\phi_{1,i})\sqrt{\rho_{i+1}\rho_{i}}+(\phi_{2,i+1}-\phi_{2,i})\sqrt{(1-\rho_{i+1})(1-\rho_{i})}\bigg]+\]\[+S_i^{\tilde{\beta}}\sqrt{\rho_{i+1}(1-\rho_{i})}e^{i(\phi_{1,i+1}-\phi_{2,i})}+S_i^{-\tilde{\beta}}\sqrt{(1-\rho_{i+1})\rho_{i}}e^{i(\phi_{2,i+1}-\phi_{1,i})}\quad\bigg).\]
We note that the coherent states at initial and final times are not integrated in the previous formula.

By the quantum mechanics formalism, the conditional probability is given by
\[
P(S(T),(n,m)|S(0),(n_1,m_1))=\]
\begin{align*}
&=\int\prod_{i,k}\frac{1}{2\pi i}d\bar{\psi}_{i,k}d\psi_{i,k}e^{-\bar{\psi}_{i,k}\psi_{i,k}}\bra{n,m}\ket{\psi_{1,k}}\bra{\bar{\psi_N}}\hat{U}(T=N\Delta ,0)\ket{\psi_0}\bra{\bar{\psi}_{2,k}}\ket{n_1,m_1}\,,
\end{align*}
where:
\[\bra{n,m}=\bra{0}\psi_{1,N}^{n}\psi_{2,N}^{m}\frac{1}{n!m!}\]
\[\ket{n_1,m_1}=\bar{\psi}_{1,0}^{n_1}\bar{\psi}_{2,0}^{m_1}\ket{0}\]\[\]
With some computations explained in $\cite{Il}$$\footnote{  pg.s. 136, 166 and 277-281}$, we  obtain the formula
\begin{align}\label{comod}
&P(S(T),(n,m)|S(0),(n_1,m_1))=\frac1{n!m!}S(T)^{-\tilde{\beta}\frac{(n-m)}{2}}S(0)^{\tilde{\beta}\frac{(n_1-m_1)}{2}}\\
&\times\int d\bar{\psi}d\psi\bra{\bar{\psi_N}}\hat{U}(T=N\Delta ,0)\ket{\psi_0}\bar{\psi}_{1,0}^{n_1}\bar{\psi}_{2,0}^{m_1}\psi_{1,N}^{n}\psi_{2,N}^{m}e^{-2M}\,,\nonumber
\end{align}
where:
\[\int d\bar{\psi}d\psi=\prod_{k=1,2}\prod_{i=0,N}\int \frac{1}{2\pi i}d\bar{\psi}_{k,i}d\psi_{k,i}\,.\]
The previous integral is expressed in the coherent state variables, which means that in order to compute it, we need to transform the whole expression in the hydrodynamical variables; Note that, as in quantum mechanics, the result of the integral is a complex number, while the probability is its module squared. We keep Ilinski's notation to simplify the comparison with his theory.

The previous formula allows us to compute the  probability density function of the final price $S(T)$ with a final portfolio allocation $(n,m)$, given the initial price $S(0)$ and the portfolio allocation $(n_1,m_1)$. Due to gauge-invariance, we can choose $S(0)=1$. This is the choice we make in all simulations.

\section{The numerical integration}

The numerical simulation  requires the computation of the approximate value of an integral in high dimension. We provide here a brief description of the strategy for the numerical integration. We first note that the integral $(\ref{comod})$ involves four variables for each time step:
\[\rho_i,\phi_{1,i},\phi_{2,i} \quad \textup{and}\quad S_i.\] The first three variables describe the orders dynamics; we denote the space of these variables the {\em hydrodynamical} space. The last variable is the stock price.\par
The algorithm first selects a particular configuration in the hydrodynamical space, which corresponds to a particular trading pattern, then it samples the associated stock price variables with the Metropolis-Hastings algorithm, which is a Markov chain Monte Carlo method, using the potential
\[\exp\bigg[-\sum_{i=0}^{N-1}\frac{1}{2\sigma^2\Delta}\bigg(\log(S_{i+1})-\log(S_{i})-2\alpha[(\rho_{i+1}-\rho_{i})]\bigg)^2\bigg].\]
The sampled values $S_i$ are then used to compute the integral $(\ref{comod})$.
Details about the Metropolis-Hastings algorithm can be found in $\cite{MeTe}$.
We recall that the following assumption have been introduced:
\begin{itemize}
	\item $\mu=r_1=r_2=0$ ,
	\item tc=0 
	\item $\tilde{\beta}$=2.5,
	\item M=$2n_{1}$=$2m_{1}$=$2n$=$2m$=100.
\end{itemize} 
The first assumption is not restrictive, and makes the results more transparent.
The second one consists in neglecting transaction costs, but we point out that these could be easily introduced in the model.
We choose $\tilde{\beta}=2.5$ as in \cite{Il}, observing that a different value of $\tilde{\beta}$ does not change the qualitative behaviour of the model. Indeed, the Hamiltonian is invariant under the transformation
$$
S_i\rightarrow S_i^{\tilde{\beta}^{-1}}\,, \sigma\rightarrow \sigma\tilde{\beta}^{-1/2}
$$
and such invariance can be used to eliminate $\tilde\beta$.

\begin{figure}[ht]
\begin{center}
		\includegraphics[width=5.5cm]{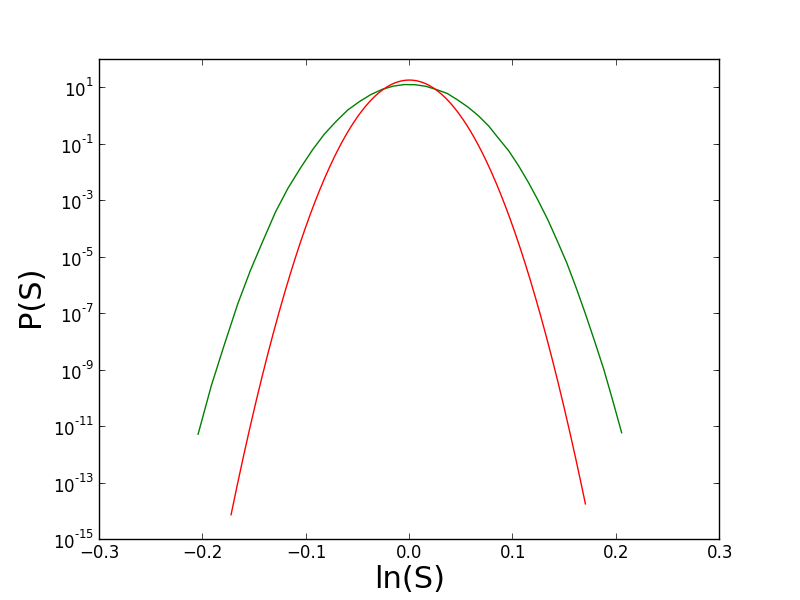}
		\end{center}
		\caption{Normal distribution (red) and PDF of the  perturbed model with $\alpha=0.461$ (green). We present only one simulation in order to show the effect of the  perturbation.} 
		\label{fsim}
\end{figure}

\begin{figure}[ht]
\begin{center}
		\includegraphics[width=5.5cm]{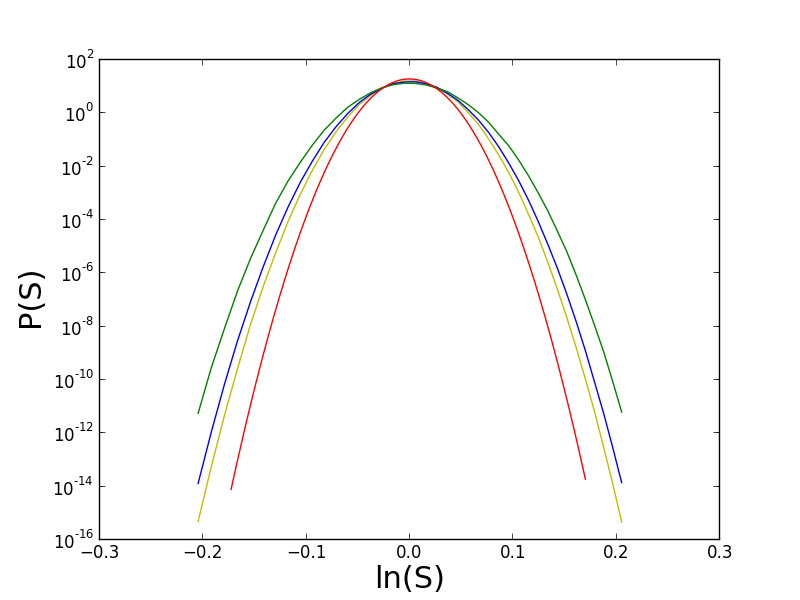}
\end{center}
		\caption{In this case all three simulations, $\alpha=0.266$ (yellow),  $\alpha=0.333$ (blue) and  $\alpha=0.461$ (green) are shown together to highlight the relationship between the perturbation intensity and $\alpha$. }
		\label{ssim}
\end{figure}

\begin{figure}[ht]
	\centering	\includegraphics[width=5.5cm]{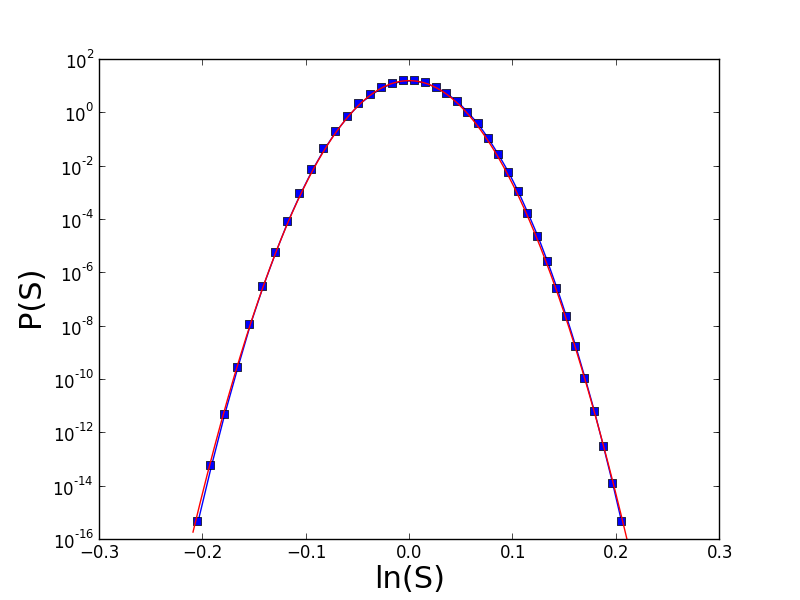}
	\caption{The red line is the normal distribution, whereas the blue line with the squares is the result of the simulation of the previous model with $\alpha=0.266$ . }
\label{figcompare}
\end{figure}

The final assumption is chosen as a compromise between the computational complexity and the accuracy;  simulations performed with higher values of $M$ show similar results.
The dynamics of the model is a perturbed Geometric Brownian Motion, the perturbation being proportional to the parameter $\alpha$.
The parameters of the Brownian Motion are the same of the simulation discussed above. The results of  simulations are shown in logarithmic scale in Figures $\ref{fsim}$ and $\ref{ssim}$.
It turns out that this perturbation only causes an increase of the variance $\sigma$. This can be observed in Figure \ref{figcompare}, where the PDF of the model without orders, with
$\alpha=0.266$ and $\sigma=0.00648$ (red continuous line), and a normal distribution with variance $1.156\sigma$ (blue line with squares) are shown.

The numerical computation of the  mean, variance and kurtosis of the simulated probability density function shows that the previous values are equal to those corresponding to the Geometric Brownian Motion  with
$\sigma=0.0074908=0.00648*1.156$, within of 0.001\%.

This poor match with financial data is also confirmed by Ilinski in \cite{Il}, where he writes that ``this strategy is far from optimal''.

\section{The generalized model}

Ilinski suggests a second kind of perturbation
\[v=-2\alpha[(\rho_{i+1}-\rho_{i})]^k-\alpha_2(\rho_{i+1}-\rho_{i})\frac{\Delta \log(S_i)}{n_1/M-1/2}\,,\]
and he computes the probability distribution of $S(T)$ with the saddle point method and other approximations in the case $k=1$.
The probability distribution displayed in \cite[p.~148, Fig.~6.15]{Il} is very accurate in the central part, but it behaves badly in the deep tails.
If we analyse the probability density function derived by Ilinski's model, we note that its wings exhibit a linear relationship between $\log(P(S(T)))$ and $\log(S(T))$.

This behaviour is not in good agreement with the stocks dynamics. The overlap between the computed and observed probability density functions is not very accurate in the wings region.
In particular we can see that the relation between $\log(P(S(T)))$ and $\log(S(T))$ showed in  the PDF is of polynomial type; i.e, 
\[
\log(P(S(T)))=\alpha \log(S(T))^{\Gamma}
\]   
We propose a different perturbation: Ilinski's action
\[-\beta\mathcal{A}_g(S)=\sum_{i=0}^{N-1}\frac{-1}{2\sigma^2\Delta}\bigg(\log(S_{i+1})-\log(S_{i})-2\alpha[(\rho_{i+1}-\rho_{i})]\bigg)^2, \]
depends linearly on the differences  $(\rho_{i+1}-\rho_{i})$. We introduce the action
\[-\beta\tilde{\mathcal{A}}_g(S)=\sum_{i=0}^{N-1}\frac{-1}{2\sigma^2\Delta}\bigg(\log(S_{i+1})-\log(S_{i})-\sum_{k=1}^{
	J}2\alpha_k(\rho_{i+1}-\rho_{i})|\rho_{i+1}-\rho_{i}|^{\Gamma_k-1}\bigg)^2,\]
where $J\ge1$ and $\Gamma^k\ge1$ are integers.

\begin{figure}[ht]
	\centering
	\includegraphics[width=5.5cm]{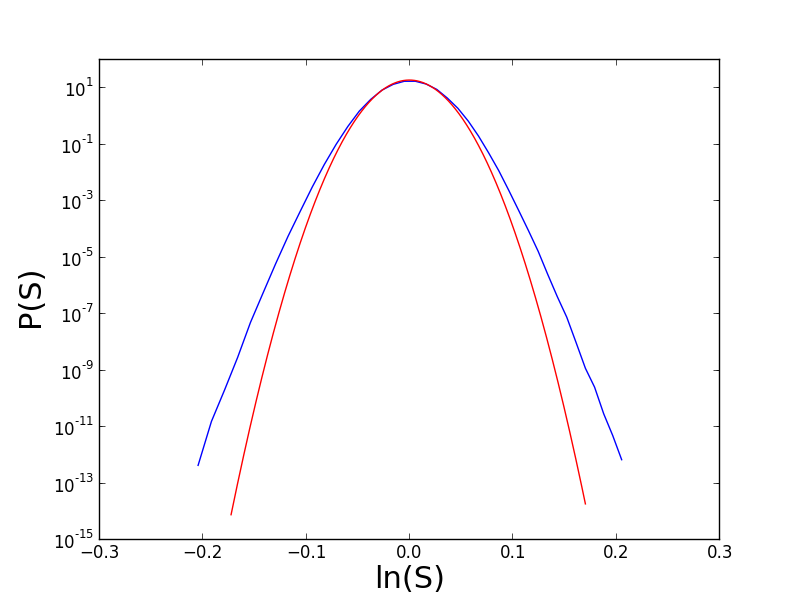}
	\caption{ The red line  is the  normal distribution, whereas  the blue line is the generalized model with
		$\alpha=0.461$.  }
\label{fig:lepto}
\end{figure}

We keep $J$ small to avoid to overfit the data; it turns out that this perturbation with $J=2$ provides results in good agreement with all the real data that we analysed.
Still, at first we present a result with $J=1$ and $\Gamma=3$ i.e. 
\[\sum_{k=1}^{J}2\alpha_k(\rho_{i+1}-\rho_{i})|\rho_{i+1}-\rho_{i}|^{\Gamma_k-1}=2\alpha(\delta\rho)^3.
\] 
The result of the numerical simulation is displayed in Figure \ref{fig:lepto}, which shows the leptokurtic behaviour
\medskip
\medskip

\begin{figure}[ht]
		\centering
		\includegraphics[width=5.5cm]{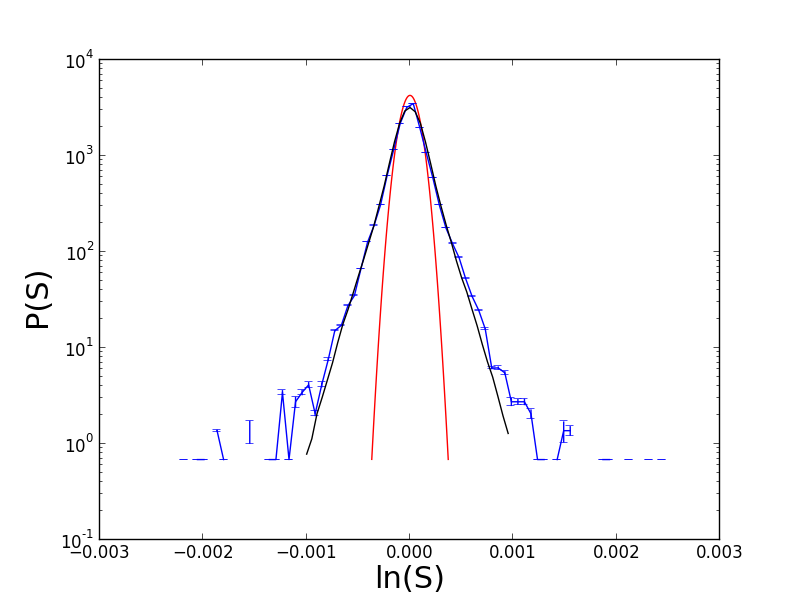}
		\caption{S$\&$P500: $\sigma = 0.0000280$, $T=10$, $\alpha= 0.00092$, $N=60$ and $\Delta \log(S)=0.002$.}
		\label{h1}
\end{figure}

\begin{figure}[ht]
\centering
\includegraphics[width=5.5cm]{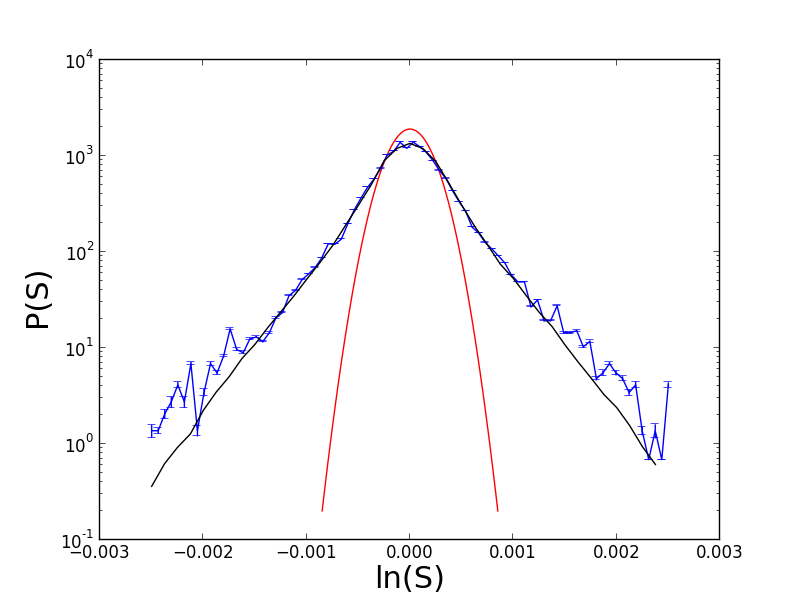}
\caption{APPL}
\small
 $\sigma = 0.0000628$, $T=10$,   $\alpha= 0.0024$, $N=80$ and $\Delta \log(S)=0.005$.
 \label{h3}
\end{figure}

We  present  some comparisons between real data and our model. The source of the real data consists in 3 months price-sheets of the  {\em S\&P500} index
and {\em APPLE} stocks from 01/05/2017 to 26/07/2017. The sampling frequency is $\tau=60 s$; each dataset consists of about 25000 prices. In order to obtain the probability density function associated to the  index and the stock, we follow the method introduced in reference \cite{Bc}, i.e. we consider a set of  historical data as instances of a stochastic variable. We first compute $X_i$ by
\[ X_i=\log(P(t_i)/P(t_{i-1})) \quad \quad t_i-t_{i-1}=\tau,  \]  
then we build a histogram of the values $X_i$ with $N$ bins. The histograms shown in Figure $\ref{h1},
%\ref{h2},
\ref{h3},\ref{h4}$ and $\ref{htail}$  display the number of counts $\Delta C$ in each histogram bin,   divided by the bin width $\Delta S/N$. The result is then normalized in order to approximate a probability density function.

The error bars for the real data  are estimated as  $ \sigma_{bins}\Delta S/N$; where  $\Delta S$ is the width of the histogram $x$-bars. In order to compare the simulations with the results obtained in \cite{Il,M-S,D-F}, we plot the probability density functions in logarithmic scale.

We plot  the results of some simulations performed with the intent to reproduce real data. Here we did not plot the simulation errors in order to keep the pictures as clear as possible; such errors have been estimated and they are within  $1\%$.

The picture show in blue the empirical distributions of S$\&$P500 %and FTSE100
index; the black curves represent the results of the simulations with the generalized model.

The red curves represent the normal distribution with the same $\sigma$ and $T$; % it is rescaled in order to improve the comparison with the simulated model.
%\begin{figure}[ht]
%		\centering
%		\includegraphics[width=5.5cm]{DefFTSE100.png}
%		\caption{FTSE100: 
%		$\sigma = 0.0000328$, $T=10$,   $\alpha= 0.001$, $N=60$ and $\Delta \log(S)=0.002$.}
%	    \label{h2}
%\end{figure}

The agreement is quite good in a very wide region near the central price \footnote{$\log(S(0))=0$.}; yet in the tails,  the model underestimates the value of the probability density function. This behaviour concerns the simulations of both the indices and the Apple stock.

This can be explained by the absence of large jumps in the simulation, i.e., our model produces a smaller amount of big price variations than the real market.
The agreement between simulations and real data can be improved with an extra term.
Heuristically, we observe that the addition of a higher degree term creates a more intense perturbation when
$|\delta\rho|\simeq 1$.
For  the {\em APPLE} share and {\em S\&P500} index two different kinds of perturbations are used:
\[
\sum_{k=1}^{J}2\alpha_k(\rho_{i+1}-\rho_{i})|\rho_{i+1}-\rho_{i}|^{\Gamma_k-1}=2\alpha_1(\delta\rho)^3+2\alpha_2(\delta\rho)^9 \quad\quad \textup{APPL},\]\[
\sum_{k=1}^{J}2\alpha_k(\rho_{i+1}-\rho_{i})|\rho_{i+1}-\rho_{i}|^{\Gamma_k-1}=2\alpha_1(\delta\rho)^3+2\alpha_2(\delta\rho)^{13} \quad\quad \textup{S\&P500.}
\]
The numerical simulations associated with the previous perturbations give the following results:

\begin{figure}[ht]
    \centering
    \includegraphics[width=5.5cm]{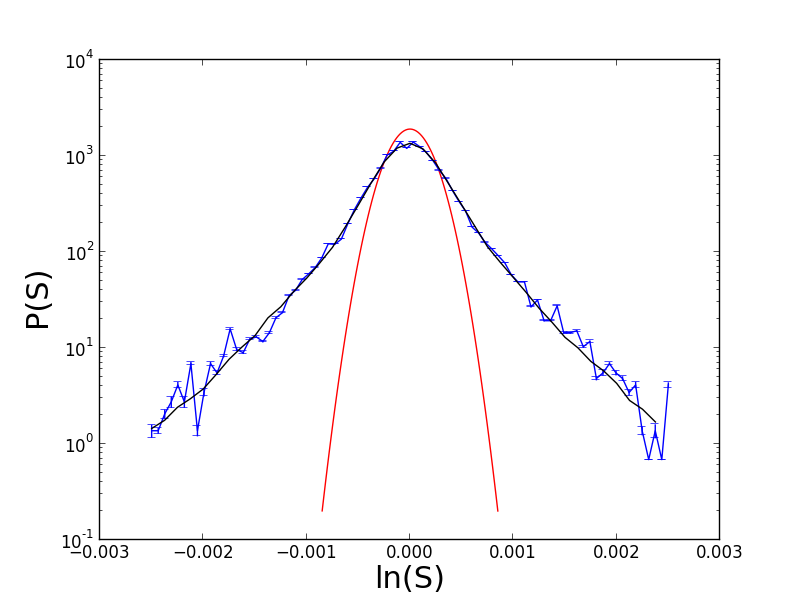}
    \caption{APPL}
    \small
       $\sigma = 0.0000628$, $T=10$,  $\alpha_1= 0.0024$, $\alpha_2= 0.0015$, $N=80$ and $\Delta \log(S)=0.005$.
       \label{h4}
\end{figure}
\begin{figure}
		\centering
		\includegraphics[width=5.5cm]{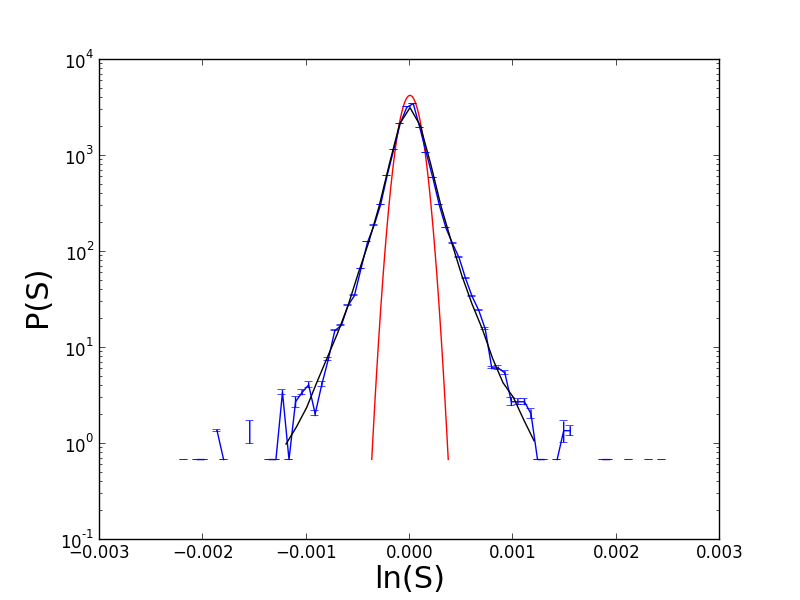}
		\caption{S\&P500}
		\small
		  $\sigma = 0.000028$, $T=10$ ,  $\alpha_1= 0.00092$ and $\alpha_2=0.0011$, $N=60$ and $\Delta \log(S)=0.0024$.
	   \label{htail}
\end{figure}

The measure that we used to quantify the agreement between real data and the simulations  is the overlap amplitude between the numerical and real data probability density functions in the y-axes. For example if the overlap spans from the point $1.0$ up to the point $2.5$ on the y-axis, we claim that the agreement is of one and half order of magnitude. We choose this particular measure because it has been used in the references we use as benchmarks.

One of the first attempt to reproduce the probability density function of a real asset with a  stochastic process is   \cite{M-S}. The authors prove that it is possible to obtain an agreement of  almost three orders of magnitude  with a L\'evy flight for the S\&P index with $\tau$= 1 min.
Better results have been achieved in \cite{D-F} by Dupoyet and  Fiebig  using a quantum lattice model which reproduces the probability density function of NSDAQ index with an agreement about four orders of magnitude with the same $\tau$. Albeit  \cite{D-F}  is a considerable improvement over   \cite{M-S}, it suffers of the same problem of  Ilinski's, that is, it underestimates the probabilities of large market corrections.

Our model fits the  APPLE stock and S\&P index probability density functions with an agreement of almost four order of magnitude, and in particular, when compared with the other models mentioned above, it provides a good fit or the distributions in the deep tails region.

\section{Tables and statistical analysis}

We provide a quantitative relationship between the statistical properties of the simulations
and the parameters values.  Two tables are presented with different values of $\alpha_1 $ and $\alpha_2$; in the first line we write the parameters associated to a GBM with the same $\sigma$ and $T$ of the generalized model.
From the tables it is possible to infer a direct relationship between the strength of the
perturbation  and the value of $\alpha_k$.
The greatest differences between the two models can be seen in the last lines of the table,
where $\alpha_1\simeq 0$, i.e.,  where the perturbation generated by $\alpha_2$ is stronger.
The estimate of the parameters is obtained by the classical formula 
\[
\textup{k-th moment}(X)=\int (x-\mu)^kf(x)dx,
\] 
where $f$ stands for the probability density function of the variable $X$ and $\mu$ is its mean.

The error in the variance, kurtosis and the other moments is estimated computing $5$ times the probability density function and then evaluating for each result the parameters. The values written are the mean and  standard deviation computed over the previous results.  
\medskip 
\begin{figure}[ht]
	
	\centering
	
	\includegraphics[width=6.5cm]{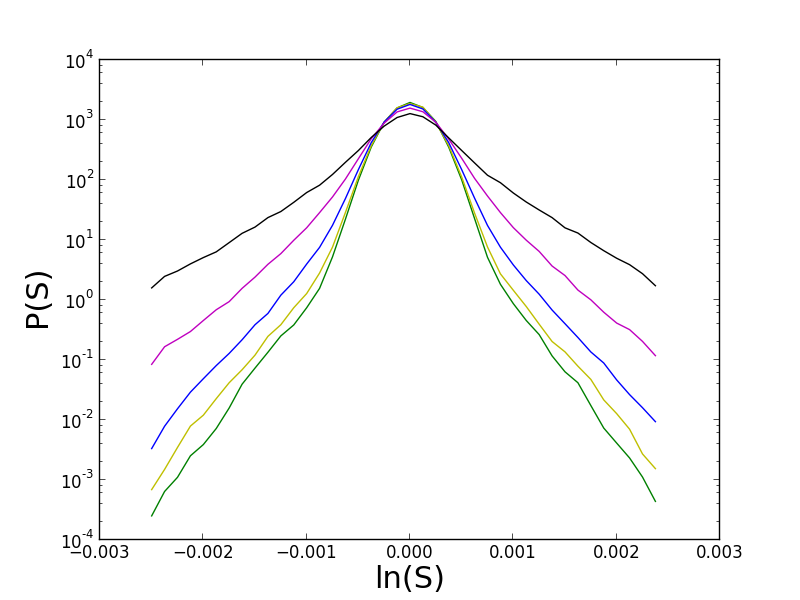}
	\small

\end{figure}
\[\]
\begin{tabular}{|l|l|l|l|r|}
	\hline
	$\alpha_{1,2}$ & Variance & Kurtosis & 6th-Moment & 8th-Moment\\
	\hline
	GBM & 3.93 e-8 & 3.00 & 9.19 e-22 & 2.53 e-28\\
	\hline
	\splitcell{$\alpha_1=2.6$ e-3\\$\alpha_2=1.1$ e-3} & \splitcell{2.0720 e-7\\ $\pm$ 1.34 e-9}  & \splitcell{6.4628\\ $\pm$ 0.017}  & \splitcell{7.6536  e-19 \\ $\pm$ 3.13 e-21} & \splitcell{2.9050  e-24 \\ $\pm$ 1.65 e-26} \\
	\hline 
	\splitcell{$\alpha_1=1.3$ e-3\\$\alpha_2=1.1$ e-3}& \splitcell[]{9.0440 e-08 \\ $\pm$  2.33 e-10} & \splitcell[]{6.2574 \\ $\pm$ 0.038} & \splitcell[]{8.8153 e-20 \\ $\pm$ 6.13 e-22} & \splitcell[]{2.5805 e-25 \\ $\pm$ 5.33 e-27}\\
	\hline 
	\splitcell{$\alpha_1=6.5$ e-4\\$\alpha_2=1.1$ e-3}& \splitcell[]{5.5517 e-08 \\ $\pm$  5.89 e-10} &\splitcell[]{5.0646\\ $\pm$ 0.065} & \splitcell[]{1.4685 e-20 \\ $\pm$ 7.80 e-24} & \splitcell[]{3.1448 e-26 \\ $\pm$ 2.88 e-28}\\
	\hline
    \splitcell{$\alpha_1=3.25$ e-4\\$\alpha_2=1.1$ e-3}& \splitcell[]{4.5458 e-08 \\ $\pm$  7.40 e-10} &\splitcell[]{4.3413\\ $\pm$ 0.015} & \splitcell[]{5.3952 e-21\\ $\pm$ 5.73 e-23} & \splitcell[]{8.6426 e-27 \\ $\pm$ 8.38 e-29}\\
	\hline
	\splitcell{$\alpha_1=1.65$ e-4\\$\alpha_2=1.1$ e-3}& \splitcell[]{4.2556 e-08 \\ $\pm$  4.41 e-10} &\splitcell[]{3.7724\\ $\pm$ 0.073} & \splitcell[]{3.4505 e-21\\ $\pm$ 3.44 e-23} & \splitcell[]{4.1386 e-27 \\ $\pm$ 4.03 e-29}\\
	\hline
\end{tabular} 
\[\]
\begin{figure}[ht]
	
	\centering
	
	\includegraphics[width=6.5cm]{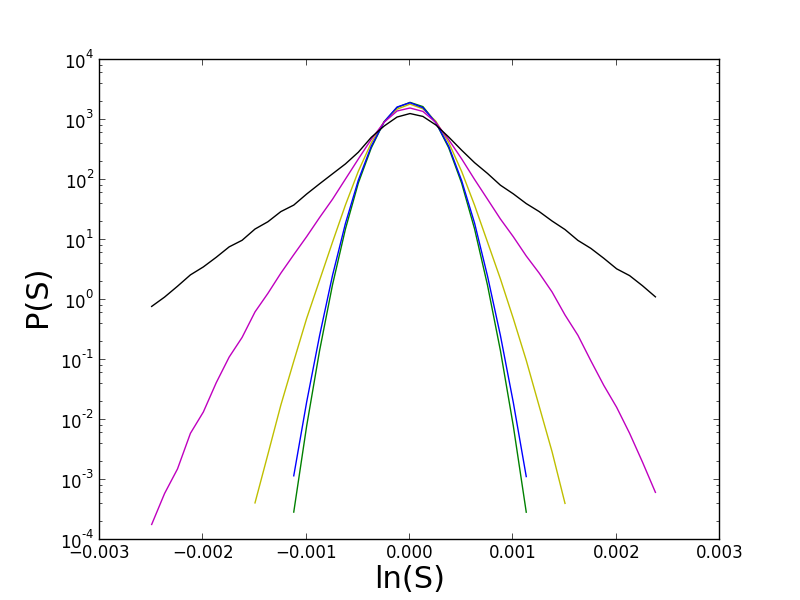}
	\small

\end{figure}
\[\]
\begin{tabular}{|l|l|l|l|r|}
	\hline
	$\alpha_{1,2}$ & Variance & Kurtosis & 6th-Moment & 8th-Moment\\
	\hline
	GBM & 3.93 e-8 & 3.00 & 9.19 e-22 & 2.53 e-28\\
	\hline
	\splitcell{$\alpha_1=2.6$ e-3\\$\alpha_2=0$} & \splitcell{1.9595 e-07 \\ $\pm$ 7.17 e-10}  & \splitcell{5.8585\\ $\pm$ 0.018}  & \splitcell{5.5543 e-19 \\ $\pm$ 1.79 e-21} & \splitcell{1.9428  e-24 \\ $\pm$ 5.05 e-27} \\
	\hline 
	\splitcell{$\alpha_1=1.3$ e-3\\$\alpha_2=0$ }& \splitcell[]{8.5264 e-08 \\ $\pm$  1.13 e-10} & \splitcell[]{4.2395 \\ $\pm$ 0.010} & \splitcell[]{2.3935 e-20 \\ $\pm$ 1.53 e-22} & \splitcell[]{3.0881 e-26 \\ $\pm$ 2.48 e-28}\\
	\hline 
	\splitcell{$\alpha_1=6.5$ e-4\\$\alpha_2=0$ }& \splitcell[]{5.0286 e-08 \\ $\pm$  4.02 e-10} &\splitcell[]{3.2557\\ $\pm$ 0.0047} & \splitcell[]{2.4980 e-21 \\ $\pm$ 1.70 e-23} & \splitcell[]{1.1181 e-27 \\ $\pm$ 3.30 e-29}\\
	\hline
	\splitcell{$\alpha_1=3.25$ e-4\\$\alpha_2=0$ }& \splitcell[]{4.4560 e-08 \\ $\pm$  2.40 e-10} &\splitcell[]{3.1126\\ $\pm$ 0.0096} & \splitcell[]{1.4168 e-21\\ $\pm$ 1.69 e-23} & \splitcell[]{4.4932 e-28 \\ $\pm$ 6.54 e-30}\\
	\hline
	\splitcell{$\alpha_1=1.65$ e-4\\$\alpha_2=0$ }& \splitcell[]{3.9401 e-08 \\ $\pm$  1.32 e-10} &\splitcell[]{3.0249\\ $\pm$ 0.012} & \splitcell[]{9.2073 e-22\\ $\pm$ 3.33 e-25} & \splitcell[]{2.6648 e-28 \\ $\pm$ 5.50 e-30}\\
	\hline
\end{tabular}
\\
\[\]
\[\]
It is also important to note that the  $\alpha_k$ values are proportional to $\sigma$. In the first generalized model simulation $\alpha/\sigma\simeq10^{-1}/10^{-3}=10^2$, which is equal to the real data cases $\alpha/\sigma\simeq10^{-3}/10^{-5}=10^2$.
This is  clear if we observe the whole action
\[\tilde{\mathcal{A}}_g(S)\simeq\bigg(\delta \log(S) -2\alpha_1(\delta\rho)^{\Gamma_1}-2\alpha_2(\delta\rho)^{\Gamma_2}\bigg)^2.\]
In the Metropolis-Hastings algorithm, in order to obtain a mixing ratio of $25\%$, the $S$ fluctuations are proportional to $\sigma$, while $\delta\rho$ is always distributed in $[-1,1]$, i.e.,
\[
\delta \log(S)\simeq\sigma,\]\[ -2\alpha_1(\delta\rho)^{\Gamma_1}-2\alpha_2(\delta\rho)^{\Gamma_2}\simeq\pm2\alpha_1\pm2\alpha_2.
\]
In order to perturb in a proper way the price variation we need
\[
\sigma\simeq \pm2\alpha_1\pm2\alpha_2\,,
\]
therefore a change in the order of magnitude of $\sigma$ must correspond to a similar change in the  parameters $\alpha_1$ and $\alpha_2$.

\section{Interpretation of the generalized model}

\begin{figure}[ht]
	
	\centering	
	\includegraphics[width=5.4cm]{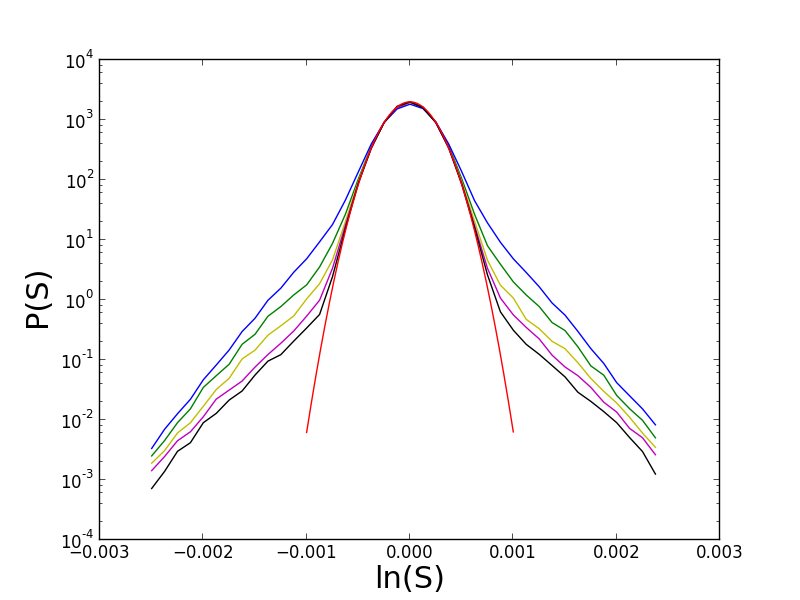}
	\caption{$\alpha(\delta\rho)^{\Gamma}$}
	\small 
	$\Gamma$=5 -blue-, 7 -green-, 9 -yellow-, 11 -purple- and 13 -black-.\\
	The red curve represents the normal distribution with the same $\sigma T$.
	\label{htail2}
\end{figure}

The terms $\alpha_k(\delta\rho)^{\Gamma_k}$ introduced above allow us to obtain a good agreement between the simulated and the real PDF.
In this section we provide a financial interpretation of  such  quantities. Given the perturbation
\[\alpha(\delta\rho)^{\Gamma}\,,\]
we observe that
\begin{itemize}
	\item $\Gamma$  affects the jumps size,
	\item $\alpha$  affects the probability of the jumps.  
\end{itemize}
\medskip
We consider $\Gamma$ first. Figure $\ref{htail2}$ shows the result of a single perturbed model  $\alpha(\delta\rho)^{\Gamma}$ with different values of $\Gamma$ and fixed $\alpha$.\\

The Geometric Brownian Motion yields a mesokurtic probability density function and all the trajectories simulated with this model have a continuous path.
However, it is possible to obtain big fluctuations between the initial price $S(0)$ and the final price $S(T)$ by setting a large variance $\sigma$. 
Increasing $\sigma$ does not affect the trajectory continuity, since the model remains a GBM.
The price fluctuations are directly linked with the variance $\sigma$; but the continuity  is not affected by the value of this parameter. These facts suggest an interpretation of $\sigma$  as the frequency of the orders with small spread $\Delta S=|S(i)-S(i+1)|$. A larger value of $\sigma$ corresponds to a larger number of orders per unit time, yielding a larger price fluctuation.  Since we are considering small $\Delta S$ variations, continuity is preserved.

This model is too simple for a real market description; in particular some massive price corrections may happen in a unit time frame. This events are called jumps.

In the real financial context, massive price corrections appear when there is an external change in the macroeconomic scenario; when this happens, the original price may be greatly underestimated or overestimated. In the previous situation the orders given immediately after the macroeconomic change  will have a large spread $\Delta S$.

Within this framework, large price corrections are more likely; which implies that the  tails of the PDF are fatter. Given  a model which allows only $x\Delta S$ jumps  with $\Delta S= |S(T)-S(0)|$,  the probability of a price change $y\Delta S$ where $y<<x$, it is equal to the case without jumps. Instead, if $y \ge x $ the probability will be much higher with respect the jumps-less case.

Denoting  as $\tilde{S}$  the  price  where the wings start to exhibit their presence, we can observe that it is proportional to   $\Gamma$. In fact we note that the overlap between the normal distribution and the black line, with $\Gamma=13$, is longer with respect to the overlap of the blue line, with $\Gamma=5$;  moreover  all the price variations y$\Delta S \ge |\tilde{S}-S(0)|$ are more likely to happen with respect to the Geometric Brownian Motion model. This is in agreement with our  interpretation on $\Gamma$.\par
We also recall that the perturbation with $\Gamma=1$ is equal to a Geometric Brownian Motion 
with increased $\sigma$; suggesting again that   $\Gamma_k$ is related with jumps size present in the model.
\medskip

\begin{figure}[ht]
	\centering	
	\includegraphics[width=5.5cm]{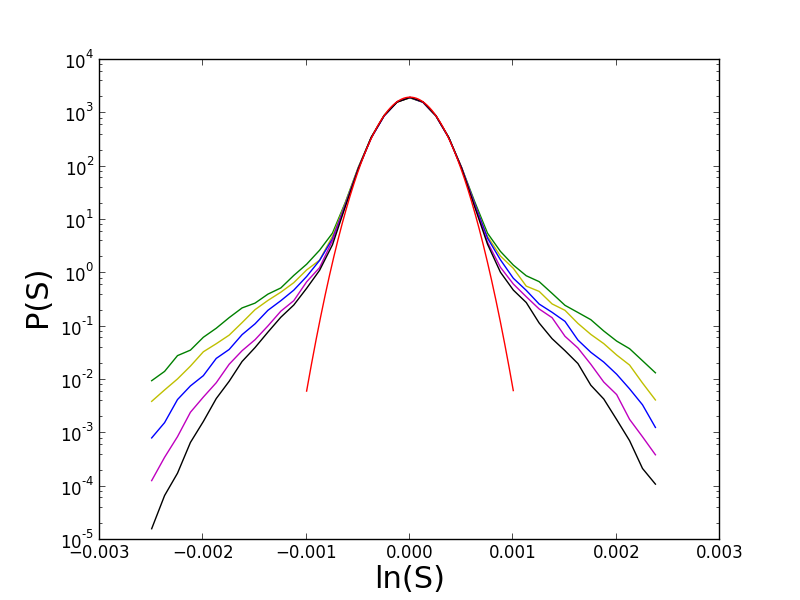}
	\caption{$\alpha(\delta\rho)^9$}
	\small 
	 $\alpha=$ 0.00181 -green-, 0.00158 -yellow-,  0.00140 -blue-, 0.00126 -purple- and 0.00114 -black-.
	 \label{slopetail}
\end{figure}

To consider the effect of $\alpha$, we show the results of a simulation with the perturbation $\alpha(\delta\rho)^9$ and different values of $\alpha$ in Figure $\ref{slopetail} $.
\[\]
The previous simulation, along with the previous tables, shows that the kurtosis and higher even moments of the distribution are directly linked with the value of $\alpha$.\par 
The mass under the tails quantify the presence of massive  price variation, occurring in presence of jumps; which means that the probability associated to this variations   is directly related with the  jumps probability. This shows the relation between jumps and the shape of the tails.

\end{document}